\documentclass[aps,prb,amssymb,amsfont,reprint,superscriptaddress]{revtex4-1}
\usepackage{amstext}
\usepackage{hyperref}
\usepackage{graphicx}
\usepackage{amstext}
\usepackage{textcomp}
\usepackage{color}
\usepackage[update,outdir=./Fig/p/]{epstopdf}

\begin{document}
\title{Hopping conduction via ionic liquid induced silicon surface states}

\author{J. \surname{Nelson}}  \email{nelson@physics.umn.edu}
\affiliation{School of Physics and Astronomy, University of Minnesota, Minneapolis, Minnesota 55455, USA}
\author{K. V. \surname{Reich}} \email{kreich@umn.edu}
\affiliation{School of Physics and Astronomy, University of Minnesota, Minneapolis, Minnesota 55455, USA}
\affiliation{Ioffe Institute, St. Petersburg, 194021, Russia}
\author{M. \surname{Sammon}} 
\affiliation{School of Physics and Astronomy, University of Minnesota, Minneapolis, Minnesota 55455, USA}
\author{B. I. \surname{Shklovskii}}  
\affiliation{School of Physics and Astronomy, University of Minnesota, Minneapolis, Minnesota 55455, USA}
\author{A. M. \surname{Goldman}}
\affiliation{School of Physics and Astronomy, University of Minnesota, Minneapolis, Minnesota 55455, USA}
\date{\today}

\begin{abstract}  
In order to clarify the physics of the gating of solids by ionic liquids (ILs) we have gated lightly doped $p$-Si, which is so well studied that it can be  called the "hydrogen atom of solid state physics" and can be used as a test bed for ionic liquids. We explore  the case where the concentration of  induced  holes at the Si surface is below  $10^{12}\text{ cm}^{-2}$, hundreds of times smaller than record values. We find that in this case an  excess negative ion binds a hole on the interface between the IL and Si becoming a surface acceptor. We study the surface   conductance of holes hopping between such nearest neighbor acceptors. Analyzing the acceptor concentration dependence of this conductivity, we find that the localization length of a hole is in reasonable agreement with our direct variational calculation of its binding energy.  The observed hopping conductivity resembles that of well studied $\text{Na}^{+}$ implanted Si MOSFETs.  
\end{abstract}

\keywords{Si}

\maketitle
\section{Introduction}
\label{sec:introduction}
In the last decade room temperature ionic liquids revolutionized the way in which weakly conducting or insulating materials are gated in order to make them metallic or superconducting. By using an ionic liquid (IL) one can achieve surface carrier densities of $n\sim10^{15}\text{ cm}^{-2}$ within the electrochemical window of the electrolyte, which is of the order of a few volts.\cite{Petach(2014),Ong(2011)}

Ionic liquids are molten salts made of negative and positive ions with relatively large radii. As a result, the Coulomb attraction energy of two adjacent opposite sign ions, $E_{C}=e^{2}/\kappa D$, is small enough so that at room temperature the salt stays molten. Here $D \simeq 1$ nm is the sum of radii of positive and negative ions and $\kappa \simeq 3$ is the dielectric constant of the IL. A negative voltage, applied between a metal coil submerged in the IL and the sample, causes negatively charged ions (anions) to migrate to the sample surface. At the sample surface an electric double layer is formed, which  can be viewed as a capacitor with nm scale separation between electrodes. The crucial advantage of IL gating is that the IL-Si junction is insulating.

So far most of the research has dealt with gating of novel materials, such as rubrene~\cite{Chris_rubrene}, underdoped YBCO ~\cite{YBCO_Goldman}, polymers~\cite{Chris_polymers}, nanocrystal arrays~\cite{Philippe_ES}, nanotubes, and graphene~\cite{IL_review}, and has been aimed at the record  induced carrier densities.

\begin{figure}
\includegraphics[width=7.3cm]{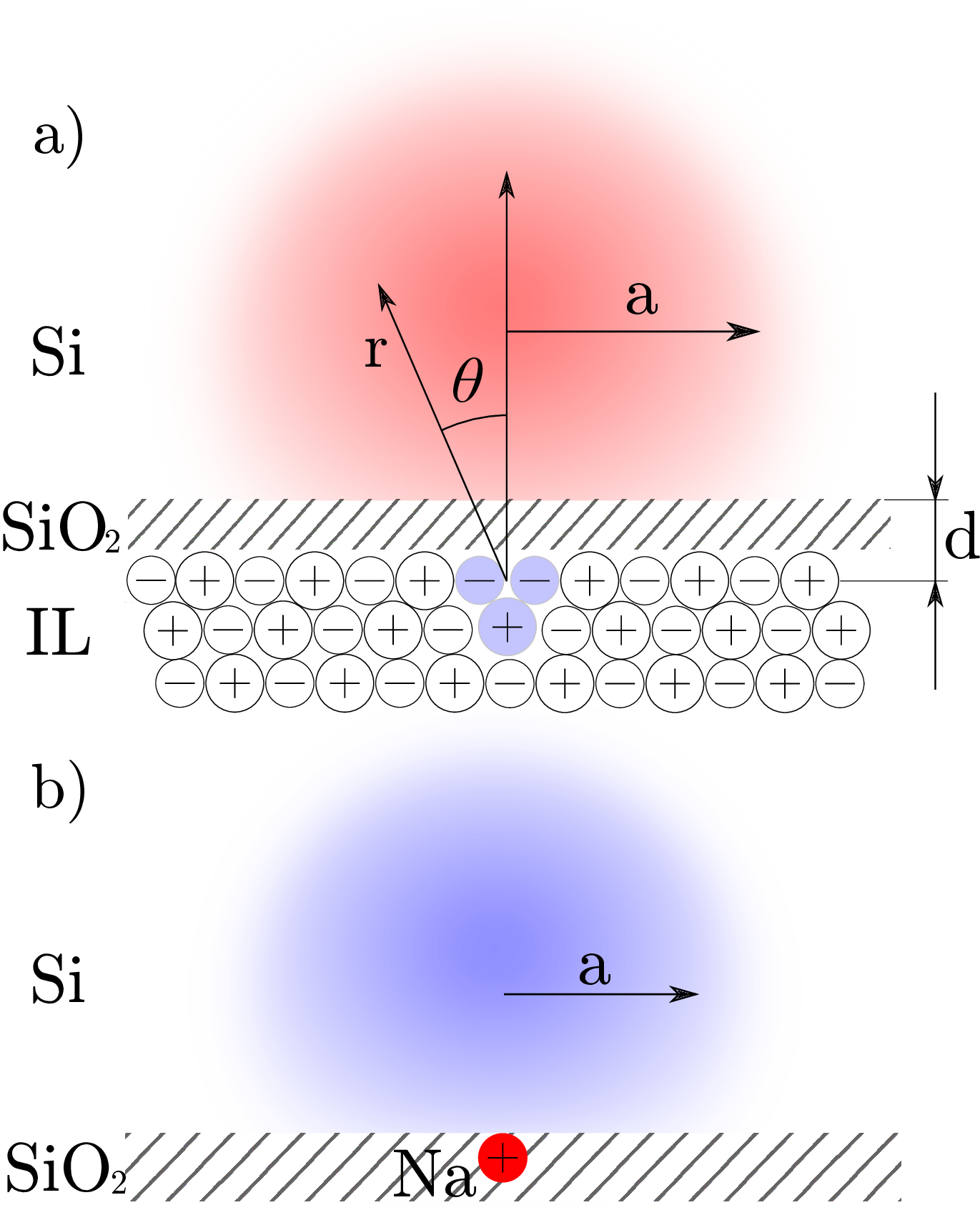} 
\protect\caption{\label{fig:pic}Surface bound states of carriers in Si (a) gated with IL and (b) gated by $\text{Na}^{+}$ implantation in the oxide. In a) the red region  represents the wave function of the hole. The negative charge acceptor cluster of 3 ions is shown by blue in a) at the distance $d$ from the Si surface. In b) the electron cloud is shown by  blue.  $a$ is the decay length of the surface bound state wave function in Si. The spherical coordinates $r,\theta$ are used below for the variational calculation of the hole binding energy.}
\end{figure}

In this paper we  focus on the basics of gating with ILs. For this purpose we choose to gate a lightly doped single crystal $p$-type Si wafer. Silicon is so well studied that it can be called ``the hydrogen atom of solid state physics'' and, therefore, used as a ``test bed'' for ILs. At large surface densities the IL charge  layer is usually regarded as an uniform sheet. In this paper we consider relatively small negative gate voltages, when the concentration of surface holes is in the range of $10^{11}\text{ cm}^{-2}<n<10^{12}\text{ cm}^{-2}$ . In this case each excess ion driven to the surface should be treated as a discrete charge individually  responsible for creation of a carrier  in a sample. 

In the absence of an applied gate voltage, the IL is a neutral and strongly correlated liquid. When $E_{C}\gg k_{B}T$ the arrangement of positive and negative ions resembles that of  NaCl. When the gate is negatively biased, additional anions are driven to the sample surface. This results in discrete excess negative ions imposed on the neutral background with NaCl-like order. An example of such a net negative group of ions at the surface of Si is shown on Fig. \ref{fig:pic}a. Below we present evidence that suggests the excess anion binds a hole residing in Si (see the hole wave function cloud in Fig. \ref{fig:pic}a.) \footnote{Previously the role of discrete excessive charges in IL was explored in the theory of double layer capacitance of IL-metal \cite{discretness_IL}}. One can notice that such a IL surface acceptor is similar to the $\mathrm{Na^+}$ surface donor in Si MOSFETs as shown in Fig. \ref{fig:pic}b.\cite{AndoRMP} 

In the $\text{Na}^{+}$ experiments, the $\text{Na}^{+}$ ions were implanted on the top of layer of $\text{SiO}_{2}$ and driven by a positive gate voltage  to the Si surface. Each $\text{Na}^{+}$ ion binds an electron near the Si surface, becoming a surface donor. The surface conductivity as a function of temperature was shown to follow

\begin{equation}
\sigma_{2D}=\sigma_{1}\exp\left(-\frac{\epsilon_{1}}{k_{B}T}\right)+\sigma_{3}\exp\left(-\frac{\epsilon_{3}}{k_{B}T}\right),\label{eq:fow}
\end{equation}
\noindent where $\epsilon_{i}$ is the activation energy and $\sigma_{i}$ the conductivity prefactor.~\cite{AndoRMP} The first term, which represents the activation of electrons to the conduction band with the activation energy $\epsilon_{1}$ equal to the ionization energy of the donor, dominates at high temperatures. The second term, observed at low temperatures, describes the nearest neighbor hopping between the $\text{Na}^{+}$ donor states at the Si surface. (Some compensation of Na$^{+}$ donors by acceptors creates empty donors and allows nearest neighbor hopping to proceed.)

In this work we show that IL gating of $p$-type Si leads to a similar transport phenomenon. We observe the nearest neighbor hopping with the activation energy $\epsilon_3$ changing with the gate voltage, while the first term of Eq. (\ref{eq:fow}) is shorted by the bulk conductance. This confirms the existence of discrete and sparse IL acceptors at the Si surface shown in Fig. \ref{fig:pic}a. 

The remainder of this paper is organized as follows. In the following Section \ref{sec:exp_details} we discuss our experimental set up. In Section \ref{sec:results} we present our experimental results and their analysis. In Section \ref{sec:model} we discuss our model of IL gating of $p$-Si with a Pt electrode and explain the origin of the negative threshold voltage. In Section~\ref{sec:parameters} we theoretically evaluate parameters of the surface acceptor and compare them with our experimental results. We conclude in Section \ref{sec:conclusion}.

\section{Experimental Details}
\label{sec:exp_details}
In our experiment we made ohmic contacts to a boron doped Si wafer after annealing it in a mixture  of $\text{N}_{2}$ and $5\%\text{ H}_{2}$ in order to reduce  the number of dangling bonds at the surface. A glass cylinder attached to the Si surface was used to confine the ionic liquid and control the gated area between contacts. The gate electrode consisted of a Pt metal coil suspended in the ionic liquid DEME-TFSI. The resistivity of the wafer is $1-5~\mathrm{\Omega cm}$ which with the typical hole mobility in Si of $400 ~\mathrm{cm^2/V s}$ corresponds to a boron concentration of $10^{16}~\mathrm{cm^{-3}}$. This concentration is smaller than 3D concentration of $10^{17}~\mathrm{cm^{-3}}$ ($10^{11}~\mathrm{cm^{-2}}/10^{-6}~\mathrm{cm}$) induced by the gate in the surface layer which has a width of $10^{-6}~\mathrm{cm}$ (see below). Additional experimental details can be found in the Supplemental Material of Ref. \onlinecite{JJ_MIT}.

Other experiments have reported the possible role of electrochemical reactions when gating  with  an IL \cite{Jeong2013,Petach(2014)}. To avoid these reactions we modulated the carrier density below room temperature at $T=230$K, where the IL ions are still mobile. A thin oxide was used to protect Si from further  oxidation and in addition, to passivate the trapping sites on the Si surface \cite{Gallagher,cobbold}. This native oxide has a thickness of $10-15\text{ \AA}$ as measured by ellipsometry (a thicker oxide would reduce the capacitance of the transistor).

The 4-terminal sheet resistance $R_{S}$ of the silicon channel was measured by the van der Pauw method. In our previous study \cite{JJ_MIT} we showed that the temperature dependence of the resistance at each gate voltage was reproducible during the cooling to 2 K and warming up to 230 K.

To calculate the surface concentration one can, in principal, use the capacitance of the device. However, in such a structure the total capacitance depends on both the geometrical and quantum capacitance \cite{Uesugi_2013}. The last one can be calculated for a degenerate gas with high hole concentration \cite{Uesugi_2013} or for   small concentrations \cite{Brian_capacitance}. But our situation is one of  intermediate concentration when the distance between holes is comparable with the Bohr radius of a hole. In this case one cannot use these formulas. Doing so would lead to uncertainties in the capacitance value. Therefore, to calculate the surface hole concentration we rely on integration of the charging current.

\begin{figure}[h]
\includegraphics[width=8.6cm]{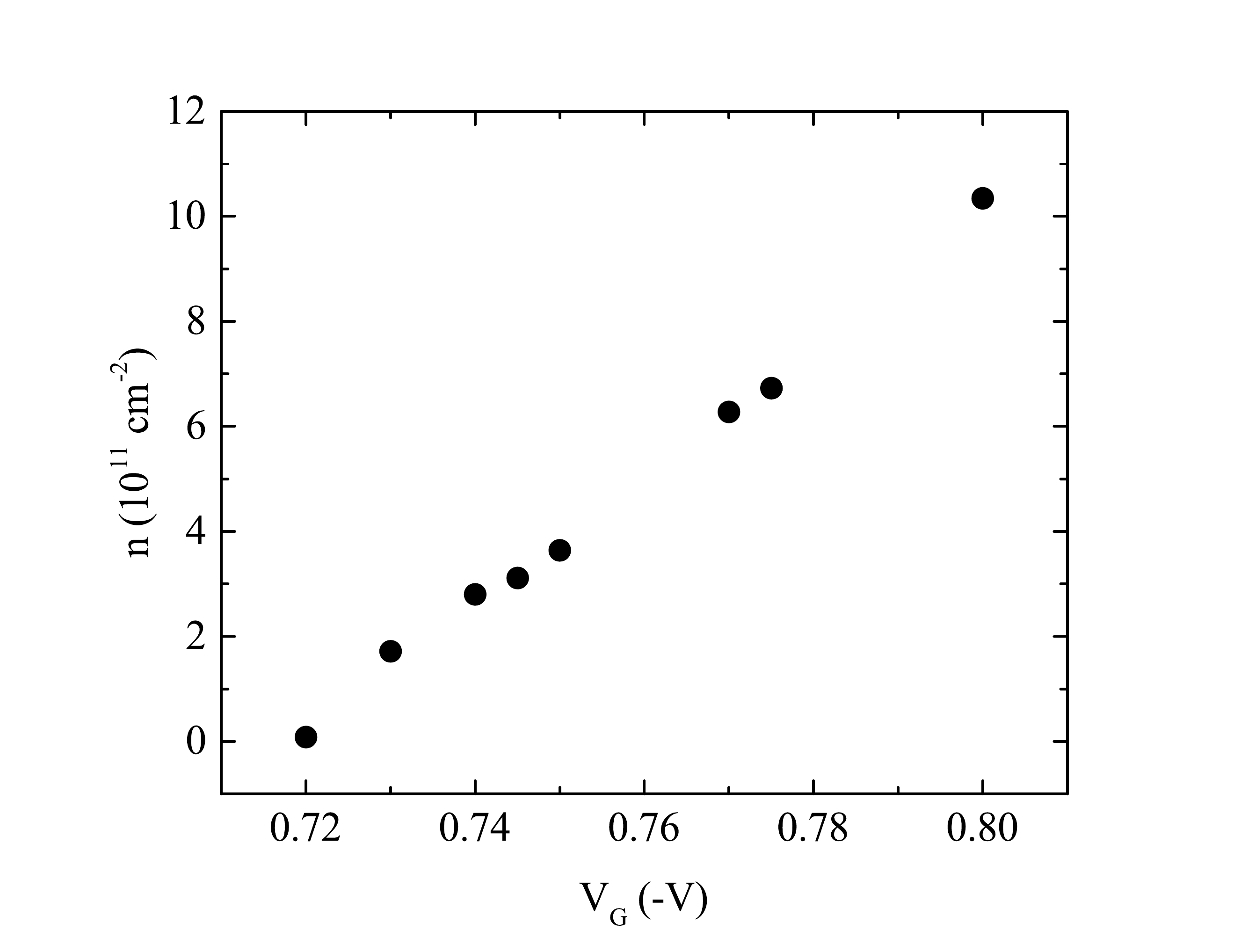}
\protect\caption[n vs Vg]{\label{fig:nV}Integrated  carrier density $n$ as a function of applied gate voltage $V_{G}$.}
\end{figure}

Just like in the $\text{Na}^{+}$ experiments we do not observe the surface conduction until a certain sample dependent threshold gate voltage $V_{T}=-0.7~\mathrm{V}$ is reached \footnote{The definition of threshold gate voltage is slightly different from the conventional one}. At a gate voltages $V_G$ lower than $V_{T}$ an additional anion brings and binds a free hole from the bulk of the Si sample, becoming a shallow acceptor. Therefore, the two-dimensional hole density, $n$, at $V_G<V_{T}$  was calculated by integrating the current during the charging process starting from $V_G=V_{T}$. The dependence of the integrated carrier density $n$ on the applied gate voltage $V_G$ \footnote{Our device is not equipped with a reference electrode to quantify the amount of applied gate voltage that actually drops across the interface between the IL and Si as it was done for example in \cite{Braga_surface, reference_electrode}. This is not an  issue as $V_G$ was not used in the calculation of the sheet carrier density.} is shown on Fig.~\ref{fig:nV}. \footnote{To study the low carrier density limit, we initially charged a sample to the conducting state and discharged it in small steps to study the low carrier density behavior. The data was taken until $V_G=V_T$ where surface state conduction vanishes. The voltage steps (100 meV) used in this experiment lead to current spikes of around 50 nA when $V_G$ was changed. Within 10 s of altering $V_G$ the current decayed back down to the noise floor of the system. While the charging process appears to only last for a few seconds the sample remained at a temperature where the IL was in a liquid state for several minutes. Waiting for longer than a few minutes had no effect on the surface state resistance.}

From the slope of the dependence of $n$ on $V_G$ one can find the capacitance of the device $2~\mathrm{\mu F / cm^2}$. Assuming that dielectric constant of the IL and the $\mathrm{SiO_2}$ is the same $\kappa_{SO}=4$ the related distance between negative ions in the IL and the holes in Si  is $17~\mathrm{\AA}$ .  This  agrees reasonably with the sum of the size of the ion and the width of the $\mathrm{SiO_2}$ layer.


Since  our device has a large area $\simeq 6 ~\mathrm{mm^2}$ and we drive relatively large current $\sim 50~\mathrm{nA}$ the noise contribution to the accumulated charge is negligible.



\section{RESULTS}
\label{sec:results}

\begin{figure}
\includegraphics[width=8.6cm]{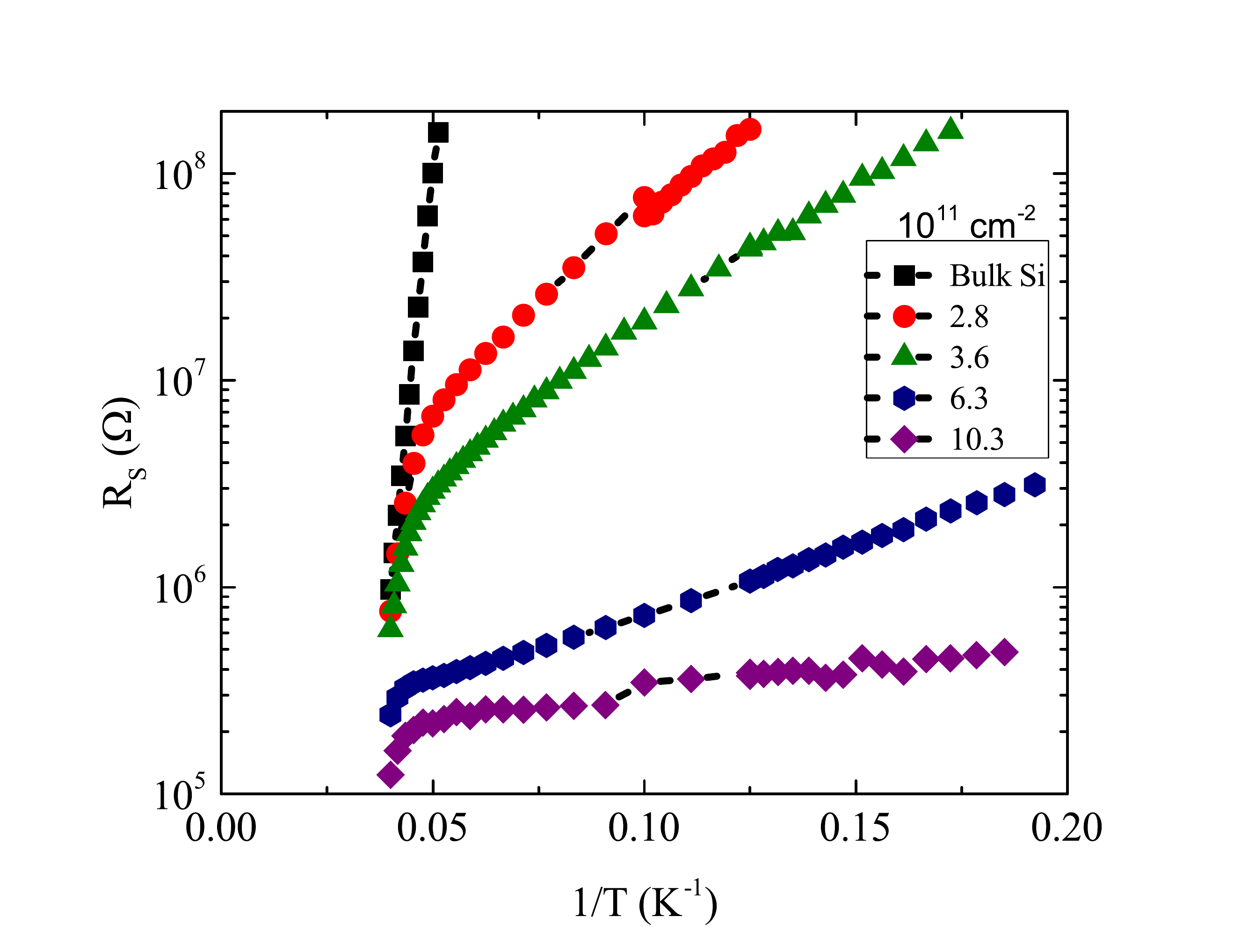} 
\protect\caption{\label{fig:RvT} Natural logarithm of the sheet resistance $R_{s}$ (in Ohms) plotted as a function of inverse temperature. Bulk conduction is independent of the gate voltage. When the ions create a surface channel, nearest neighbor hopping is observed at low temperatures. The four curves correspond to different surface concentrations $n$, in units $10^{11}~\mathrm{cm^{-2}}$. Two carrier densities are omitted for clairity.}
\end{figure}

Plotted as a function of inverse temperature, the sheet resistance $R_{S}$ is shown in a logarithmic scale in Fig. \ref{fig:RvT} at different $V_G$ corresponding to four different concentrations of holes. Above $20\text{ K}$, $R_{S}$ plotted in this way follows a straight line independent of $V_G$

\begin{equation}
R_{S}=R_0 \exp\left(\frac{\epsilon_{b}}{k_{B}T}\right),\label{eq:bulk}
\end{equation}

\noindent where $\epsilon_{b}=43\text{ meV}$ and $R_0$ is a prefactor. The ionization energy of a Boron acceptor in Si is known to be $45~ \text{meV}$, so that it is natural to assume that this line corresponds to the bulk conductance. Below $20\text{ K}$, $R_{S}$ continues to follow activated temperature behavior, but with a $V_G$ dependent activation energy. 

It is natural to interpret this low temperature part of Fig. \ref{fig:RvT} as the nearest neighbor hopping conductivity of holes between IL acceptors, which can be described by the second term of Eq. (\ref{eq:fow}).

In our $p$-type Si samples $1/R_0 \gg\sigma_{1}$; therefore, the $\sigma_{1}$ part of the surface conductivity of Eq. (\ref{eq:fow}) is shorted by the bulk conductance. This shorting  does not  occurs in $\text{Na}^{+}$ doped samples because  they are in the inversion layer regime, where the conducting path between source and drain electrode is insulated from the bulk of the sample by the hole depletion layer, while in our case we deal with an accumulation layer.

\begin{figure}
\includegraphics[width=8.6cm]{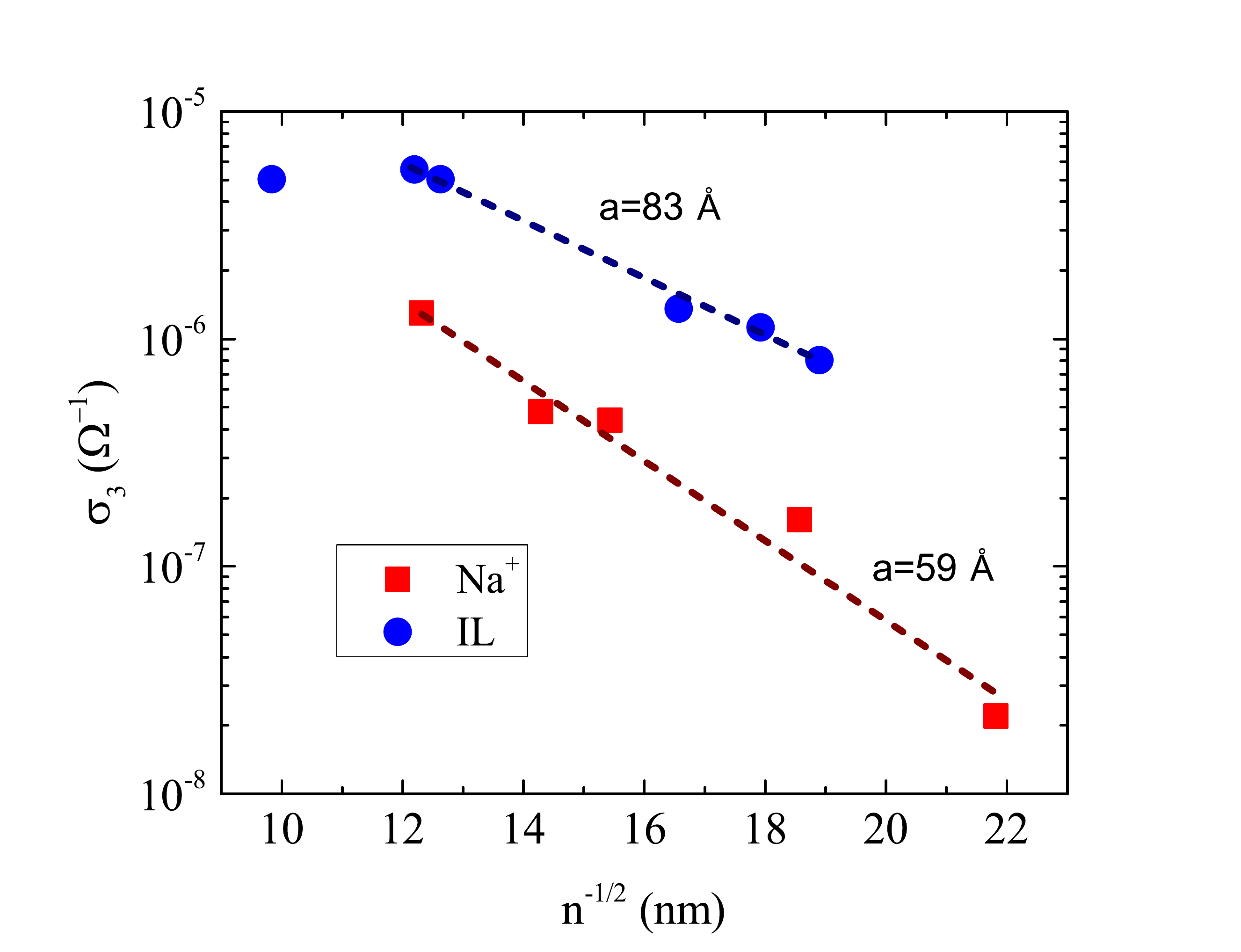} 
\protect\caption{Prefactor of nearest neighbor hopping conductivity vs. the inverse square root of the carrier density of surface carriers induced by IL gating (this work) and by Na$^{+}$ implantation (Ref. \onlinecite{AndoRMP}). Straight line fits are drawn to evaluate the decay length $a$ of the localized carrier wave function on IL acceptors and $\text{Na}^{+}$ donors as explained in the text. The size of the symbols reflects experimental uncertainty. }
\label{r3} 
\end{figure}

From each plot of Fig. \ref{fig:RvT} we extracted the prefactor $\sigma_{3}$ and the activation energy $\epsilon_{3}$ which are shown in Fig. \ref{r3} and Fig. \ref{fig:e3} respectively, as functions of the carrier density $n$. For comparison we added the corresponding data for the nearest neighbor hopping conduction of $\text{Na}^{+}$ implanted Si MOSFETs as summarized in Table III of Ref. \onlinecite{AndoRMP}. Remarkably, the two experiments cover the same range of carrier concentrations and show similar hopping parameters.

The dependence $\sigma_{3}(n)$ can be compared with the theoretical prediction based on the percolation approach \cite{percolation_1,Pollak1972486,percolation_impurity_band,percolation_magnetic_fields} as explained in Chapters 5 and 6 of Ref. \onlinecite{ESbook}. The two-dimensional analog of Eq. (6.1.9) with the help of Table 5.3 gives

\begin{equation}
\sigma_{3}=\sigma_{30}\exp[-2.4/\sqrt{na^{2}}],\label{sig3}
\end{equation}

\noindent where $\sigma_{30}$ is a conductivity prefactor. In Fig. \ref{r3} we  compare our data with Eq. \ref{sig3}. A reasonably good fit of the data is provided by the two straight lines corresponding to this equation. For the decay lengths we get $a=59 \pm 10 \text{\AA}$ for Na and  $a=83 \pm 20\text{\AA}$ for IL.  The relatively large error in $a$ for the IL case comes from the accuracy of our carrier density which was determined from the uncertainty in $V_T$ to be $\Delta n  = 6 \times 10^{10}~\mathrm{cm^{-2}}$.

\section{Model of ionic liquid gating}
\label{sec:model}
Let us present our model of IL gating in greater detail than in the introduction. Let us discuss what happens when a Pt electrode gates $p$-Si crystal with help of an IL but still at $V_G=0$. The work function of Pt,  $\varphi_{Pt}\simeq 5.7~\mathrm{eV}$, while the work function of $p$-Si, $\varphi_{Si}=\chi + E_g=5.1~\mathrm{eV}$, where $\chi=4~\mathrm{eV}$ is  the electron affinity and  $E_g=1.1~\mathrm{eV}$ is the band gap of Si. In equilibrium the electrochemical potentials of Pt and $p$-Si have to be equal. This means that electrons are transfered from $p$-Si to Pt. This results in  a positively charged $p$-Si and a negatively charged Pt. The resulting electric field between them is compensated by the IL polarization, so that anions move to $p$-Si and cations move to the Pt surface. The voltage difference $\varphi_{Pt}-\varphi_{Si}\simeq 0.6 ~\mathrm{eV} $  drops at the capacitance of the IL/Si surface, because the area of the Pt electrode is much larger than that of the Si one. Thus, there is an excess number of anions  on the Si surface even in an equilibrium. This means that  ideal contact Pt/IL/Si accumulates a layer of holes in Si. In such a situation, the  Si wafer surface would be a good $p$-type conductor. 

In fact the situation is different. An ungated $p$-type Si wafer is covered by a thin oxide layer. It is known that the oxide-silicon interface contains a substantial concentration of oxygen vacancies, which play the role of surface donors in $p$-Si~\cite{Deal}.  This results in an inversion layer of electrons on the ungated surface of   $p$-Si. In other words, the Si bands are bent downwards by the value of the Si gap  $E_g$ creating a pocket for electrons while holes are depleted in a wide layer near the surface.

Let us see what happens when such a real surface is gated by an IL with a  Pt electrode. In this case, $p$-Si donates some of its electrons to Pt and becomes positive. As result of the IL polarization an equal number of anions  move close to the Si surface forming a capacitor with positive donors. Actually at the temperature 230 K where gating is done the Coulomb attraction energy of an anion to a positive donor is substantially larger than $k_BT$ so that most of the anions form compact pairs with empty positively charged donors. The remaining donors still create a downward bending of the conduction and valence bands and the depletion layer of holes, but this bending is smaller than $E_g$ and does not create an inversion layer near the surface. Thus, at $V_G = 0$ the Si surface is an insulator.

To make it conducting we apply the negative $V_G = V_T \simeq - 0.7 ~\mathrm{eV}$. This threshold voltage brings enough anions to the Si surface to neutralize all the remaining donors. When all  the donors are paired with anions, the electric field applied to Si and the downward bending of the Si the bands near the Si surface vanish, making the bands flat. At $V_G < V_T$ excessive anions bend the bands near the surface upwards and form a hole accumulation layer. Each excess anion binds a Si hole forming a surface acceptor discussed in the Introduction  and shown in Fig. \ref{fig:pic}. These acceptors  overlap weakly with each other. At the same time their effective three-dimensional density easily exceeds the density of the bulk B impurities so that at low temperatures the nearest neighbor hopping of holes between surface acceptors dominates the sample conductance (see Fig. \ref{fig:RvT}).  

Recall that surface acceptors are situated on a background of many compact pairs of donors and anions. These pairs are compact dipoles and create only a relatively weak random potential. In the following discussion of  nearest hopping transport of holes on surface acceptors we ignore the potential of typical pairs. However, a small fraction of these pairs plays an important role in nearest neighbor hopping. These are pairs, which are thermally dissociated, i. e. where donors have their partner anion lost to the bulk of the IL. Such ionized donors play the role of a compensating impurity for the nearest neighbor hopping. They are responsible for the small number of empty surface acceptors, which allow holes to move along chains of surface acceptors (see Chapter 3 of Ref. 19).

\section{Binding energy and the hole decay length of the surface acceptor}
\label{sec:parameters}
In this section we theoretically calculate the binding energy and the decay length $a$ for an isolated IL acceptor shown in Fig. \ref{fig:pic}a. Let us first review the acceptor theory in the bulk $p$-Si. It is known that the  valence band is doubly degenerate and has two types of holes, light and heavy,  with masses $m_{l}=0.16~m_{e}$, $m_{h}=0.46~m_{e}$ respectively, here $m_{e}$ is the electron mass. It is known \cite{Diakonov} that for the above ratio $m_{h}/m_{l}\simeq2.9$ the ground state energy of the acceptor is $E_{0}=me^{4}/(2\kappa_{Si}^{2}\hbar^{2})$, where $m=0.7~m_{h}=0.3~m_{e}$ and $\kappa_{Si}=12$ is the dielectric constant of Si. If we assume that a bulk acceptor can be treated as a hydrogen atom with a hole mass $m$, then the wave function of the first excited $p$-state  vanishes at the acceptor location and has a binding energy $E_{0}/4$ \cite{acceptor_states_in_GaAs}.

Let us now switch to the IL acceptor. If the distance between the acceptor nucleus and Si surface, $d$, is so small that $d\ll a$, one can assume, in the zero approximation, that the hole is also located just near the interface. Then the interaction between the hole and acceptor can be described by an effective dielectric constant $\kappa=(\kappa_{Si}+\kappa_{SO})/2=8$,  where $\kappa_{SO}=4$ is the dielectric constant of $\text{SiO}_{2}$. (For simplicity we make the approximation that dielectric constant of the IL is that of $\text{SiO}_{2}$.) The interface between $\text{SiO}_{2}$ and Si is treated as an infinite barrier, so that the wave function of the hole vanishes at the interface. The ground state for a hole in the half space is the first excited state for a hole in bulk with a corrected dielectric constant $\kappa$, i.e. $\epsilon_{1}=Ry/4=16~\mathrm{meV}$, where $Ry=me^{4}/(2\kappa^{2}\hbar^{2})=63~\mathrm{meV}$. At the same time, the decay length $a$ is determined by the light mass

\begin{equation}
a=\frac{\hbar}{\sqrt{2m_{l}\epsilon_1}}\label{eq:size_wave_function}
\end{equation}
For $\epsilon_{1}=16~\mathrm{meV}$ we get $a=38\text{\AA}$.

Above we did not take into account that a hole interacts with its image charge and that $d$ is not zero. To do so, we use a variational approach with a probe hole wave function

\begin{equation}
\label{eq:Psi}
\Psi=2 \frac{(r\cos\theta-d)}{r_0^2 \sqrt{\pi(d+2r_0)}}\exp\left(\frac{d-r}{r_{0}}\right).
\end{equation}
Here we use spherical coordinates shown in Fig. \ref{fig:pic}a, $r>d$ is the distance between the acceptor and  the hole, $\theta$ is the polar angle and the distance $r_{0}$ minimizes the energy: 
\begin{equation}
\epsilon_{1}=\int\Psi H\Psi dV.\label{eq:energy_var}
\end{equation}
Here the integration is over the Si half space, $\theta$ ranges from 0 to $\pi/2$ and $r$ ranges from $d/\cos\theta$ to $\infty.$ The Hamiltonian for the problem is 
\begin{eqnarray}
H=-\frac{\hbar^{2}}{2m}\nabla^{2} & - & \frac{2e^{2}}{(\kappa_{Si}+\kappa_{SO})r}\label{eq:Hamiltonian}\\
 & + & \frac{e^{2}}{4(r\cos\theta-d)}\frac{\kappa_{Si}-\kappa_{SO}}{\kappa_{Si}(\kappa_{Si}+\kappa_{SO})}.\nonumber 
\end{eqnarray}
Here the first term represents the kinetic energy of a hole in Si. The second term is the potential energy of interaction of the hole  with the acceptor . The third term is the interaction between the hole and its image in $\text{SiO}_{2}$. As a rough estimate, we choose $d$ as the sum of the width of $\text{SiO}_{2}$ layer, $10\text{\AA}$ and the radius of the negative ions in the IL (TFSI) $4.4\text{\AA}$ to be $d\simeq15\text{\AA}$. In the result, for $d=15\text{\AA}$ we get that $\epsilon_{1}=8~\mathrm{meV}$ and $a=53\text{\AA}$. This decay length for a single acceptor can be considered as a reasonable estimate from below to the experimental $a=83\text{\AA}$. The concentrations of carriers which we work with are far from the case of the  light doping when the separation between acceptors is much larger than $a$. The  wave function overlap of neighboring acceptors leads to an additional increase of the  experimental value of  $a$. 

\begin{figure}[h]
\includegraphics[width=8.6cm]{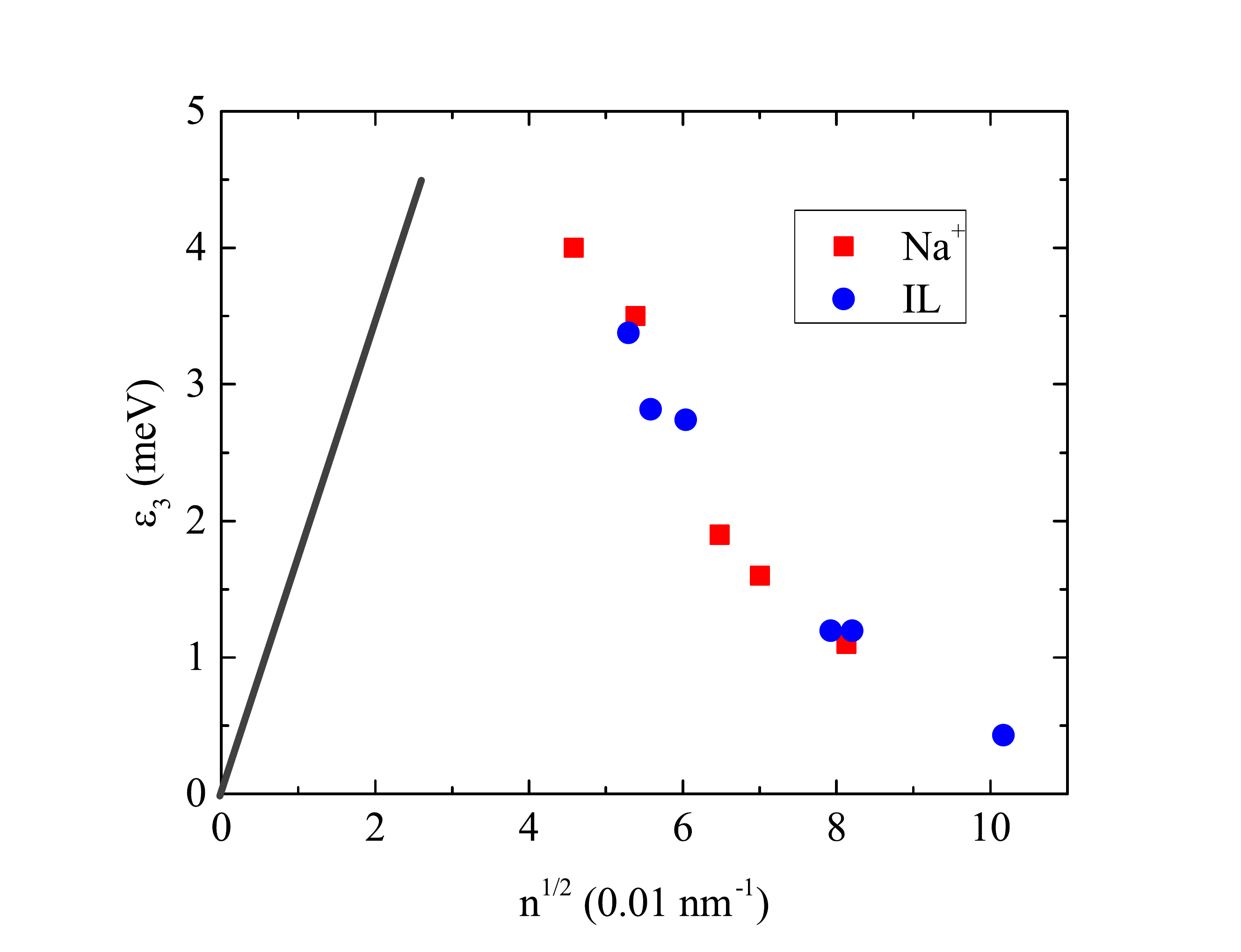} 
\protect\caption{\label{fig:e3} Activation energy of hopping conductivity of surface carriers induced by IL gating and $\text{Na}^{+}$ implantation (Ref. \onlinecite{AndoRMP}) as a function of the square root of the carrier density. The sizes of symbols reflect experimental uncertainty. The solid line following Eq.~(\ref{eq:E_3_classical}) represents the predicted dependence of $\epsilon_{3}$ on $n$ at small $n$ .}
\end{figure}

The fact that the overlap between acceptor states is substantial is also supported by Fig. \ref{fig:e3}. Here the straight line follows the equation:

\begin{equation}
\epsilon_{3}=0.98\frac{e^{2}}{\kappa}n^{1/2},\label{eq:E_3_classical}
\end{equation}

\noindent which we derived for a two-dimensional lightly doped and weakly compensated semiconductor in a manner similar to the derivation for three-dimensional case in Ref. \onlinecite{3D_classic_E_3} (see also Chapter 3 in Ref. \onlinecite{ESbook}).  Equation (\ref{eq:E_3_classical}) takes into account only classical shifts of levels of holes  induced by the Coulomb potential of rare free of IL anions surface donors (see end of previous section \ref{sec:model}) and  ignores the overlap of hole wave functions. We see in Fig. \ref{fig:e3} that in contrast to Eq. \ref{eq:E_3_classical}  the experimental $\epsilon_{3}$ values decrease with the growing  $n$. This happens because our concentrations are too large . At even larger concentrations $n>10^{12} ~\mathrm{cm^{-2}}$ the activation energy $\epsilon_3$ vanishes and the system goes through the metal-insulator transition studied in the Ref. \onlinecite{JJ_MIT}. In three-dimensional semiconductors a larger range of concentrations was studied, which displayed both regimes of growing and decreasing $\epsilon_{3}$ (see Fig. 8.1 in Ref. \onlinecite{ESbook}).

The ground state energy of Na$^{+}$ donors near the surface of Si was estimated in Ref. \onlinecite{AndoRMP}. This case is complicated by an additional electric field in the inversion layer which tends to localize electrons near the interface $\text{Si/SiO}_{2}$, to increase the donor ionization energy $\epsilon_{1}$ and to decrease the decay length $a$ of the wave function of a single donor. For experimental values of the electric field, the energy $\epsilon_{1}=35$ meV was obtained in Ref. \onlinecite{Na_theory,AndoRMP}. Using Eq. (\ref{eq:size_wave_function}) and electron light mass $m_{l}\simeq0.2m_{e}$ we arrive at $a=23\text{\AA}$. Again, due to the high experimental concentration $n$, this value can be considered to be a reasonable estimate from below for the experimental result $a=59\text{\AA}$.

\section{Conclusion}
\label{sec:conclusion}
We have explored the surface conductivity of Si gated by an ionic liquid. We studied small induced concentrations of holes $n$ below $10^{12}~\mathrm{cm^{-2}}$ and showed that at such concentrations the transport is due to the nearest neighbor hopping between acceptor like states  formed by sparse excessive negative ions of the IL. This transport is similar to the nearest neighbor hopping  between Na$^{+}$ donors in Na$^{+}$ implanted Si MOSFETs. The similarity is not only qualitative but also quantitative due to the similar average masses of holes and electrons in Si. This brings us to the conclusion that an ionic liquid acts as a set of sparse acceptors on the surface of Si.

In the future it would be interesting to extend these results to smaller hole densities, where our theory should work better. One can study different ILs and see how the sizes of  ions and  dielectric constant of ILs affect hopping conductivity. One can also extend this work to new challenging materials using the same IL so that contrasting  results 
with Si will help to understand  better the parameters of these materials. For example, the extension of measurements of surface conductivity in IL gated rubrene \cite{Chris_rubrene} to lower hole  concentrations may help to understand nature and parameters of charge carriers.

\begin{acknowledgments}
We are grateful to H. Fu for careful reading of the manuscript and  C. Leighton and B. Skinner for helpful discussion. J. N. and A. M. G. were supported by the National Science Foundation (NSF) under award DMR-1263316. K. V. R. and B. I. S. were supported primarily by the National Science Foundation through the University of Minnesota MRSEC under Award Number DMR-1420013. Devices were fabricated at the Minnesota Nanofabrication Center which receives partial support from the NSF through the NNIN program. 
\end{acknowledgments}

\end{document}